\documentclass{article}
 \PassOptionsToPackage{numbers, compress}{natbib}

 \usepackage[preprint]{neurips_2020}

\usepackage[utf8]{inputenc} 
\usepackage[T1]{fontenc}    
\usepackage{hyperref}       
\usepackage{url}            
\usepackage{booktabs}       
\usepackage{natbib,graphicx}
\usepackage{amsfonts}       
\usepackage{amssymb}
\usepackage{nicefrac}       
\usepackage{microtype}      
\usepackage{cancel}
\usepackage{soul}
\usepackage{color}
\usepackage{mathrsfs}
\usepackage{booktabs}
\usepackage{appendix}
\title{Unstructured Primary Outcome in Randomized Controlled Trials}

\author{%
  Daniel Taylor-Rodriguez\\
  
  Fariborz Maseeh Department of Mathematics and Statistics\\
  Portland State University\\
  Portland, OR 97201 \\
   \AND
   David Lovitz \\
   Fariborz Maseeh Department of Mathematics and Statistics\\
  Portland State University\\
  Portland, OR 97201 \\
   \AND
   Nora Mattek \\
   OHSU Oregon Center for Aging and Technology (ORCATECH) \\
   Oregon Health \& Science University \\
   Portland, OR 97239\\
   \AND
   Chao-Yi Wu \\
   OHSU Oregon Center for Aging and Technology (ORCATECH) \\
   Oregon Health \& Science University \\
   Portland, OR 97239\\
   \AND
   Hiroko Dodge \\
   OHSU Oregon Center for Aging and Technology (ORCATECH) \\
   Oregon Health \& Science University \\
   Portland, OR 97239\\
   \AND
   Jeffrey Kaye \\
   OHSU Oregon Center for Aging and Technology (ORCATECH) \\
   Oregon Health \& Science University \\
   Portland, OR 97239\\
   \AND
   Bruno M. Jedynak \\
   Fariborz Maseeh Department of Mathematics and Statistics\\
  Portland State University\\
  Portland, OR 97201 \\
   \texttt{bruno.jedynak@pdx.edu} \\
  
}

\begin{document}

\maketitle

\begin{abstract}
 The primary outcome of Randomized clinical Trials (RCTs) are typically dichotomous, continuous, multivariate continuous, or time-to-event. However, what if this outcome is unstructured, e.g., a list of variables of mixed types, longitudinal sequences, images, audio recordings, etc. When the outcome is unstructured it is unclear how to assess RCT success and how to compute sample size. We show that kernel methods offer natural extensions to traditional biostatistics methods. We demonstrate our approach with the measurements of computer usage in a cohort of aging participants, some of which will become cognitively impaired. Simulations as well as a real data experiment show the superiority of the proposed approach  compared to the standard in this situation: generalized mixed effect models.   
 
\end{abstract}
\section{Introduction}
Lower back pain (LBP) represents the leading cause of years lived with disability globally, ranking first in both developed and developing countries, see \cite{vos2016global}. How to test whether patients with back pain will benefit more from, say acupuncture, compared to acupressure? one of the difficulties in conducting a trial is that there is no consensus on the primary outcome. Actually, there would possibly be consensus in case the primary outcome would combine physical functioning, pain intensity, and health-related quality of life.  However, how do we assess RCT success and compute sample size with a high dimensional primary outcomes? 

A second example is provided by holistic medicine. The Academy of Integrative Health \& Medicine define health as the conscious pursuit of the highest level of functioning and balance of the physical, environmental, mental, emotional, social and spiritual aspects of human experience, resulting in a dynamic state of being fully alive. One of the challenges of organizing a RCT in holistic medicine is to define an acceptable primary outcome. It is likely that any consensus on the primary outcome will lead to a high dimensional primary outcome.  

For still another example, consider the case of aging subjects living in homes with embedded technology, measuring their sleep patterns, computer usage, walking speed, and more, see e.g. \cite{favela2015living}. One might be interested in a treatment involving regular video-conferencing for increased social interactions, see \cite{dodge2015web}. How to ``best" quantify the efficacy of this treatment in slowing the progression of cognitive impairment? The current approach consists in bi-yearly hospital visits where cognitive tests are performed. However, the frequency with which these tests are conducted makes them unsuitable to detect subtle changes in cognitive status, resulting in prohibitively long, large and expensive trials. Conversely, having technology embedded in people's homes provides the opportunity to collect high frequency digital markers, which could potentially be much more sensitive to a treatment. However, 
how to design a RCT with high frequency digital markers as primary outcome? 

In this paper, we explore the use of kernel methods for designing a simple two-arm RCT when the primary outcome is a high dimensional or an otherwise complex data type, including sequences, images, audio recordings, or wearable sensors data.  

\section{Method}
\begin{figure}[h]
\centering
\includegraphics[scale=0.37]{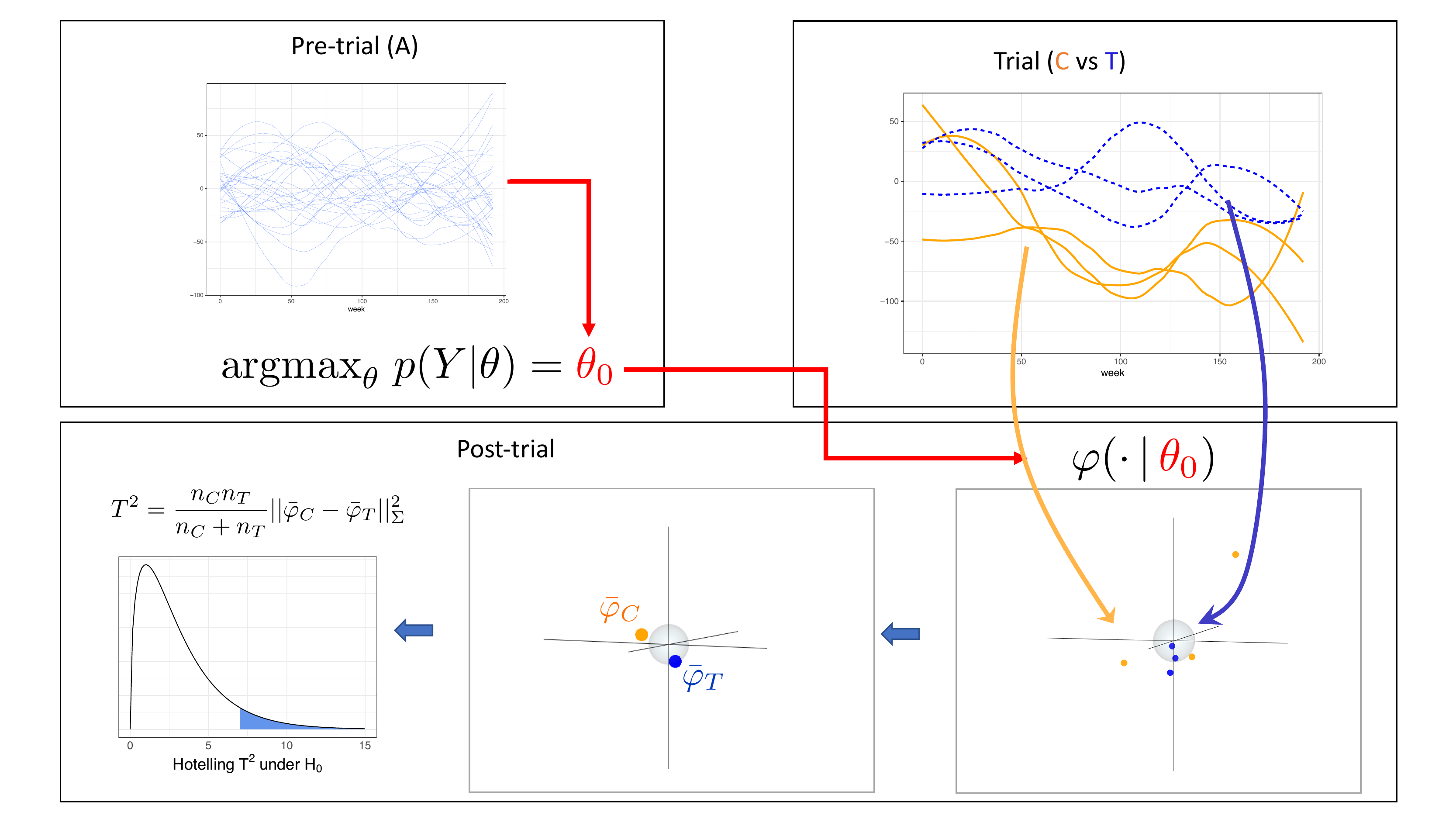}
\caption{Schematic representation of Kernel testing approach.}
\label{fig:scheme}
\end{figure}

Kernel methods are popular methods in machine learning, see e.g. \cite{scholkopf2002learning,kung2014kernel}. Recent developments in kernel methods allow for performing two sample tests of equality of distributions using Maximum Mean Discrepancy, see \cite{gretton2012kernel} and the kernalized Hotelling's test \cite{eric2008testing,harchaoui2013kernel},  Even if more work is needed, kernel methods allow for a natural generalization of the classical design of a 2-arms randomized clinical trials (RCT), proceeding as follows:  
\begin{enumerate}
    \item[Step A]  Map the data aka measurements into a {\em feature space}. The feature space is a Hilbert space, specifically a reproducing kernel Hilbert space, see \cite{scholkopf2002learning}. This mapping is established by choosing a kernel function;
    \item[Step B] Choose a univariate test statistic in feature space using a geometric construction;
    \item[Step C] Compute the finite sample or otherwise asymptotic distribution of the test statistic under the null hypothesis. Use this to threshold the test statistic; 
    \item[Step D] Establish a local alternative using another geometric construction in feature space; 
    \item[Step E] Compute the finite sample or otherwise asymptotic distribution of the test statistic under the local alternative. 
    \item[Step F] Use preliminary data, or a pilot study for computing the effect size. Combine with the previous step for computing power and sample size;  
\end{enumerate}
The classical construction of RCTs is included in this program. It is the special case where the feature space is the same as the measurement space. However, accommodating for unstructured primary outcome requires the more general construction presented. Figure \ref{fig:scheme} shows a schematic representation of a specific {\em kernel method} approach for determining the outcome of a RCT. It is an example of this program. Here, data-point are longitudinal sequences and the kernel used is the Fisher kernel. Alternatives are discussed in the discussion section. 

The schematic in Figure \ref{fig:scheme} reads as follows: before the RCT, during the pre-trial phase,  data is collected from {\em asymptomatic} (A) subjects (upper left of figure \ref{fig:scheme}). These subjects determine the healthy baseline for the trial. We then define a family of models for the data, here a Gaussian processes with a parameterized mean and covariance operator. We fit this model to the A subjects, obtaining a parameter vector $\theta_0$. The Fisher kernel is then used to map the measaurements to a Euclidean space of the same dimension as the space of parameters, concluding Step A. The data is then collected for the RCT, as shown on the upper right side. After the trial ends,  {\em Fisher vectors} are computed for each subjects' sequence of measurements (shown on the bottom right), mapping each measaurement of the trial into the feature space. The empirical estimate for the mean elements of both the treatment and control groups are computed (bottom middle). The outcome of the trial is decided using the Hotelling statistic (bottom left) (Step B).  The power is computed under a local hypothesis. Steps (C,D,E, and F). 
The details of the method are presented in the case of arbitrary data in the Appendix \ref{sec:AA}. 

\section{Results}
\subsection{Simulations}
We demonstrate the method described in this paper on an Alzheimer's data set from the Oregon Center for Aging and Technology (ORCATECH). 

\begin{figure}[h]
\centering
\includegraphics[scale=0.47]{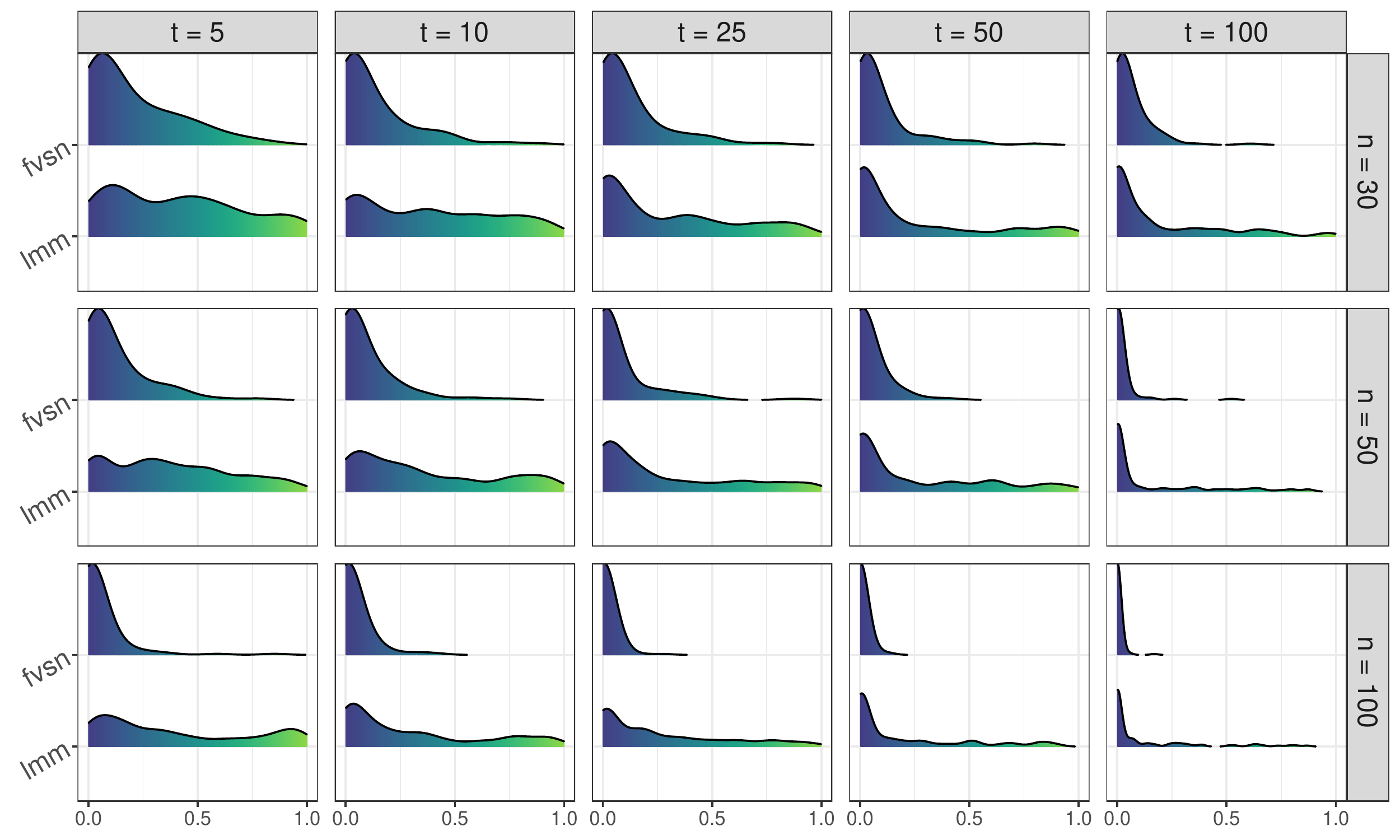}
\caption{FvH: Fisher vector Hotelling test, and lmm: linear mixed effect model. n: number of subjects. t: number of time-points. Note that when the error type I is u (e.g. .05), the power is the integral of this distribution on the left side of u (e.g. .05).}
\label{fig:sims}
\end{figure}

In order to determine the ability of the proposed approach to correctly identify differences between treatment and control groups we ran a simulation experiment using the strategy described in Figure \ref{fig:scheme}.  Specifically, we first generate data for control group, as well as for treated group  
using the Gaussian process model defined in appendix \ref{sec:AD}. The parameters were fit using the A group in the actual dataset. In generating the treated subjects, we choose the same parameters, aside for the parameter $\mu$, see Appendix \ref{sec:AD}.  In these experiments we evaluated how the power changes as a function of the sample size, and of the frequency (number of time points within a fixed total trial time) at which the measurements are collected.  Results from this simulation experiment are displayed in Figure \ref{fig:sims}, which provides the estimated sampling distribution of the p-values when the alternative hypothesis of effective treatment is true. In each case, the power is the integral of the shown distribution on the left hand side of the $\alpha$ cutoff, e,g, $\alpha=5\%$. We see that the proposed approach (fvsn) provides better power than the standard approach: the linear mixed effect model (lmm) in all cases presented, thus allowing for smaller sample size when planning for a RCT. Note also that measurements collected at higher frequencies (from left to right in figure \ref{fig:sims}) systematically improve the power for both methods.     

\subsection{Computer usage data} 
\label{sec:cu}
The ORCATECH data set which was used included 86 cognitively normal (CN) subjects and 11 subjects who developped mild cognitive impairment (MCI) during the data collection period. We considered 150 weeks of  computer usage chronological sequences. Details about the preprocessing is presented in appendix \ref{sec:AE}. Figure \ref{fig:examples-trajectories} show a few examples of trajectories. 

Using the ORCATECH data set we can simulate a RCT using cross validation. First we divide the 86 CN subjects into 8 groups of 11 (two subjects are repeated). For each group we now have 11 "treated" (CN), 11 "control" (MCI), and 75 healthy out of sample (CN). We can then test the Fisher Kernel technique's ability to determine the outcome of the 8 simulated RCTs.

To compare the behavior of the proposed approach to more traditional methods, we contrast our results from a power analysis to those obtained from the linear mixed model. The details of this approach are presented in Appendix \ref{sec:AC}.   

For this power analysis we considered the alternative distribution given in (\ref{eq:pow}). The results obtained with the kernel method yield more stable results in this small data example, needing for all data folds for the CN less than 175 observations to achieve a power of 80\%.

\begin{figure}[h]
\centering
\includegraphics[scale=0.4, trim=0cm 4cm 0cm 1cm]{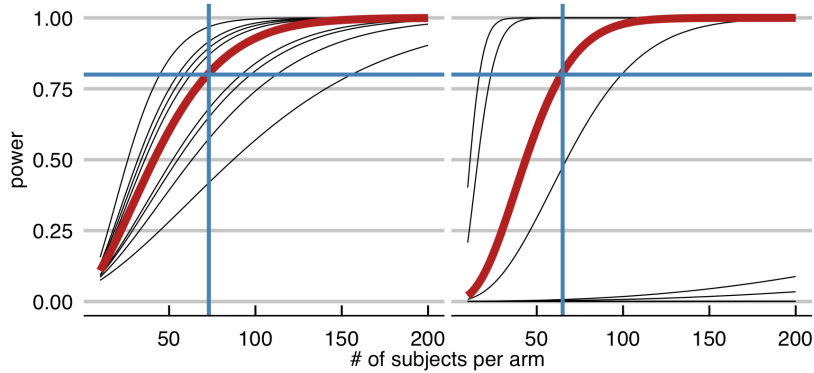}
\caption{Power vs sample size using the Fisher vector approach (left) and the linear mixed model (right). Power curves in black obtained using estimates using data from each of the 8 CN groups considered against the MCI. Red curves correspond to the power curves using the average parameter estimates from all 8 groups.}
\label{fig:power}
\end{figure}
\section{Discussion and broader impact}
We have presented a program for designing a randomized control trial when the primary outcome is unstructured. We have shown that kernel methods offer a natural generalization of the current techniques used by bio-statisticians. A key highlights are as follows: 
\begin{enumerate}
    \item A group of {\em asymptomatic} as well as a group of {\em symptomatic} subjects is used to compute the mapping into feature space, or equivalently, to compute the kernel during the pre-trial phase. These groups are also used to compute the sample size. Note that a pilot study is always available during a pre-trial phase providing the group of symptomatic subjects. The novelty here is the explicit use of a group of asymptomatic subjects in the design of the trial. 
    \item The power of the trial is computed using a geometric interpretation: the treatment allows to {\em translate} the -features of - treated group in the direction of the -features of - the asymptomatic group; 
\end{enumerate}
This last statement bring ethical questions. It could be argued that using this program, a treatment, or drug,  could be deemed successful even if no medical landmark has been crossed. On the other side, some would argue that a computational and data-based notion of health encapsulated in the definition of a kernel provides a reasonable surrogate when no simple primary outcome is available. There are also technical questions arising for this work. In particular, it is still a research topic to compute power for local alternative with the MMD or the kernel Hotelling's test.     

\section{Appendix A: Details of the method}
\label{sec:AA}
\subsection{Kernel methods and the feature space}
To define the problem more formally, we now set some notation. Let $\mathcal{X}$ denote the set of measurements that will be acquired during a RCT. To work with a concrete example in mind, let us think of the measurements as longitudinal sequences with missing values spread across. There is no structure provided with $\mathcal{X}$; no natural way to add up elements of $\mathcal{X}$ for example. Instead, we assume that we are provided with a mechanism to compare two elements $x,x' \in \mathcal{X}$. Specifically, we assume that the similarity between $x$ and $x'$ is defined by a positive definite kernel $K(x,x')$. In section \ref{sec:kernel}, we will discuss options to estimate a kernel using a dataset of healthy subjects, but until then, let us assume that this kernel is given to us. One way to visualize the action of a  kernel is to think of it as a mapping $\phi$ from $\mathcal{X}$ to a vector space $H$, also called the {\em feature space}, or in other words, each observation $x \in \mathcal{X}$ is mapped to a vector $\phi_x$. 
\begin{equation}
\label{eq:phi}
    x \in \mathcal{X} \mapsto \phi_x \in H. 
\end{equation}
During a two-armed RCT, the subjects in both groups (treatment and control) are mapped onto vectors in the feature space. If vectors from the two groups are clustered together, the trial is unsuccessful. However, if it is not the case, should the trial be consider successful? Note that in the case of a low dimensional outcome,  think of  blood pressure (2 dimensions), medical science allows to define the range of measurements with are consistent with an asymptomatic subject, eventually correcting for the effect of age, sex, and other covariates. In the case of unstructured primary outcome, there is no such science available to define a beneficiary treatment. Instead, we propose to use a {\em dataset of asymptomatic subjects}. Then, if the treatment allows to geometrically translate the feature vector of the treated subject in the direction of the feature vector of the asymptomatic group, the trial will be successful, pending it is significant.  

The fact that the kernel is positive definite lends more structure to the feature space $H$. It provides an inner product such that norms, orthogonal projection, and other simple geometric constructions can be obtained. In some cases the mapping described above is explicit and can be computed for each observation. In other cases it is implicit and only the kernel $K$ is provided. In the former case, most quantities of interest can still be computed, using the ``kernel trick", see \cite{scholkopf2002learning}. Note that the set $H$ can be infinite dimensional, specifically a set of functions. This explains why we use $\phi_x$ instead of $\phi(x)$, which allows us to denote by $\phi_x(y)$ the $y^{th}$ coordinate of $\phi_x$, both in the finite and infinite dimensional cases. 

Coming back to the technical aspects of the RCT setting, how to measure the separation between the two groups of vectors? There is not a single way to do this. However, there are simple geometric constructions that provide alternative answers. A few of these will be reviewed in Section \ref{sec:test}. In all cases, a univariate positive test statistic, notated, $U^2$ is computed. The trial is then successful if and only if the tests statistic exceeds some predetermined threshold $u_0^2$. Computing the threshold as well as estimating the power of the test requires computing probabilities in $H$ which we discuss next.          
\subsection{Mean and covariance operators} 
Consider a probability measure $\mathbb{P}$ defined over $\mathcal{X}$. For example, consider a finite set of observations $x_1,\ldots,x_n$ and consider the empirical distribution over these observations. Now, map this probability in feature space using (\ref{eq:phi}). Does this provide a probability distribution over $H$? It doesn't because $H$ is much larger, and thus most elements of $H$ have no weight! However, $\mathbb{P}$ allows to define the {\it mean element} $\mu \in H$, by
\begin{equation}
    \mu = \mathbb{E}[\phi_X]=E[K(X,.)], \mbox{ where } X \sim \mathbb{P}.
\end{equation}
Surprisingly, even though it is a single point in $H$, the mean carries a lot of information about $\mathbb{P}$. If the Kernel is {\em characteristic}, see e.g. \cite{fukumizu2009kernel},  all the information about $\mathbb{P}$ is encapsulated in $\mu$ and revealed by computing inner products. For any $f \in H$, 
\begin{equation}
\mathbb{E}[f(X)] = \langle \mu,f \rangle_H
\end{equation}

If the mean element allows us to compute the expected values of functions of a random element $X \in \mathcal{X}$, what about the covariance between two functions $f$ and $g$ of a random element $X \in \mathcal{X}$? This can be computed using the {\it covariance operator}, which is the linear operator $\Sigma_X: H \to H$ defined by
\begin{equation}
    Cov[f(X),g(X)]=\langle f,\Sigma_X g\rangle_H
\end{equation}

\subsection{Randomized clinical trials}
We consider a simple two armed RCT. In the treatment arm, we observe $\vec{x}_T=(x_1,\ldots,x_{n_T})$, while in the control, or untreated arm, we observe $\vec{x}_C=(y_1,\ldots,y_{n_C})$. Each of the $n_T+n_C$ observations belong to $\mathcal{X}$. $\vec{x}_T$ and $\vec{x}_C$ are random samples from the respective probabilities $\mathbb{P}_T$ and $\mathbb{P}_C$ over $\mathcal{X}$. Assessing the success or failure: $H_0: \mathbb{P}_T=\mathbb{P}_C$ of the trial consists in calculating a statistic $U^2(\vec x_T,\vec x_C)$ and comparing it to a threshold $u_0^2$. This threshold is such that 
\begin{equation}
    P_{H_0}(U^2>u_0^2)=\alpha
\end{equation}
for a preset value $\alpha$, say $\alpha=5\%$.  The trial is a success when the actual value $u^2$ observed satisfies $u^2>u_0^2$. The performance of the test procedure is characterized by the power of the test 
\begin{equation}
    \beta=P_{H_a}(U^2>u^2)
\end{equation}
where $H_a$ is a specific alternative hypothesis.  
\subsection{Test statistics for kernel methods}
\label{sec:test}
We present two test statistics that have been introduced for kernel methods. 
\subsubsection{Maximum Mean Discrepancy}
Recall that a kernel maps each data point in $\mathcal{X}$ to a vector in $H$. A natural idea for measuring how the two arms differ is as follows: 
\begin{enumerate}
    \item Map each data point in the treatment arm to the feature space $H$ and compute the mean vector in feature space $\hat \mu_T$;
    \item Map each data point in the control arm to the feature space $H$ and compute the corresponding mean vector $\hat \mu_C$;
    \item Define the test statistic 
    \begin{equation}
        \tilde{U}_{MMD}^2=||\hat \mu_T-\hat \mu_C||_H^2
    \end{equation}
\end{enumerate}
It is not difficult to see that $\hat \mu_T$ and $\hat \mu_C$ are the mean elements associated with the empirical probability over respectively $\vec{x}_T$ and $\vec{x}_C$. In practice, an unbiased version of $\tilde{U}_{MMD}^2$ is used, specifically, using the kernel trick,  
\begin{equation}
    U_{MMD}^2=\frac{1}{n_T(n_T-1)}\sum_{i \not = j}K(x_i,x_j) +\frac{1}{n_C(n_C-1)}\sum_{i \not = j}K(y_i,y_j)-\frac{2}{n_T n_C}\sum_{i,j}K(x_i,y_j) 
\end{equation}

How to compute the threshold $u_0$?  one can use the asymptotic distribution of $U_{MMD}^2$ under $H_0:\mathbb{P}_T=\mathbb{P}_C$ when both $n_T$ and $n_C$ are large, use a concentration inequality, or, alternatively, use a permutation test. For the later, having observed $\vec{x}_T$ and $\vec{x}_C$, assign randomly the labels $T$ and $C$ to the $n_T+n_C$ samples, while maintaining constant the number of samples in each group and compute $U_{MMD}^2$. Then, repeat this sampling procedure a large number $m$ of times independently provides $m$ - dependent - samples of $U_{MMD}^2$, under the null distribution.  $u_0$ is then the $(1-\alpha)$ quantile of the empirical distribution of these samples. Computing power for a local alternative for the MMD is challenging. Instead, we rely on the Kernel Hotelling test for which computing an approximate power, when the dimension of the feature space is small, is easier. 

\subsubsection{Kernel Hotelling test}
The kernel Hotelling test statistic is 
\begin{equation}
\label{eq:kh}
    U_{KH}^2 = \frac{1}{\sqrt{2d_2}}\left( \langle \hat \mu_C- \hat \mu_T,\left(\hat \Sigma_W + \gamma I\right)^{-1}(\hat \mu_C-\hat \mu_T)\rangle - d_1\right)
\end{equation}
where $\gamma(n_T,n_C)>0$ is a regularization constant, $\hat \Sigma_W$ is the covariance operator associated with the mixture empirical distribution  
\begin{equation}
    \mathbb{P}_W = \frac{n_T-1}{n_T+n_C+2}\mathbb{P}_T + \frac{n_C+1}{n_T+n_C+2}\mathbb{P}_C 
\end{equation}
and $d_1$ and $d_2$ are constants used to studentize the distribution, that is, insuring that the asymptotic distribution of $U_{KH}^2$ under $H_0$
 would not depend on $(\mathbb{P}_T,\mathbb{P}_C)$
 
 \begin{equation}
    d_1 = Tr\left(\left(\hat \Sigma_W + \gamma I\right)^{-1}\hat \Sigma_W\right) 
\end{equation}
and 
\begin{equation}
    d_2 = Tr\left(\left(\hat \Sigma_W + \gamma I\right)^{-2}\hat \Sigma_W^2\right) 
\end{equation}
The kernel Hotelling statistic generalized the Hotelling statistic to infinite dimensional feature spaces. However, if the trace of the operator $\Sigma_W$ is not too large, or if the feature space is of finite and small dimension, one can choose $\gamma=0$ and use the distributional properties of the Hotelling statistic. This path is taken below in the experiments. However, a more refined analysis might be needed otherwise. Also, note that there are alternatives to (\ref{eq:kh}). In \cite{li2020adaptable}, the Regularized Hotelling’s T2 defined in \cite{chen2011regularized} is used and the authors provide a data-driven selection mechanism for the regularization parameter based on maximizing power under local alternatives.   
\subsection{Power and sample size with the Hotelling statistic}
Computing power and sample size for a RCT requires specifying a single hypothesis $H_a \in H_A$. 

As discussed in the introduction, we assume that we have access to  historical data of asymptomatic subjects (A) as well as symptomatic subjects (S). From these two dataset, we compute $\hat \mu_A$ and $ \hat \mu_{S}$, the respective empirical mean elements. We then compute the power at 
\begin{equation}
  H_a:  \mu_T - \mu_C = (1-\rho)(\hat \mu_{A} -  \hat \mu_{S}) \label{eq:pow}
\end{equation}
In words, a successful treatment moves the distribution of the participants in a direction parallel to the segment connecting the asymptomatic and symptomatic subjects and with proportional magnitude $1-\rho$.    
A typical value for $\rho$ is $\rho=40\%$. A pilot study would allow us to compute a treatment specific value for $\rho$. 

The classical finite dimensional distributional results for the Hotelling test allowing to compute power and sample size are presented in the appendix \ref{sec:AB}.  

\subsection{Generative models and Fisher Kernels}
A key elements for kernel methods to be effective is to choose, and/or estimate from data an appropriate kernel. While there are standard techniques for selecting a kernel, we are interested in data driven techniques. Specifically, we assume that we dispose of one set of observations from $n_A$ asymptomatic subjects $\vec z = (z_1,\ldots, z_{n_A})$. Note that even if data might be available for symptomatic subjects, it makes sense from a statistical perspective to perform modeling only for asymptomatic subjects which likely form a more homogeneous group. Note that we will still need a sample from the symptomatic group in order to compute sample size but it is not needed for estimating the kernel.  

We now describe the Fisher kernel technique applied to this setting. Alternative techniques include auto-encoders \cite{goodfellow2016deep}, variational auto-encoders \cite{doersch2016tutorial} and generative adversarial networks \cite{goodfellow2014generative}. We choose a parametric family of probabilities $\mathbb{P}_\theta$ over $\mathcal{X}$, with $\theta \in \mathbb{R}^d$. Notate the density or point mass function $f(x,\theta)$, $x \in \mathcal{X}$. The parameter $\theta$ is then estimated using the data $\vec{z}$ through maximum likelihood, penalized likelihood or a Bayesian inference. Using this estimate $\hat\theta$ we can then calculate the Fisher information matrix $I(\hat \theta)$. The Fisher vector is a function of both $I(\hat \theta)$ and of the Fisher score.  The Fisher score evaluated at a data-point $x \in \mathcal{X}$ is defined as the gradient of the log density at that point 
\begin{equation}
     \phi(x) = \nabla_\theta \ln f(x,\theta=\hat \theta).
\end{equation}
The Fisher vector of a data-point $x$ at $\theta$ is 
\begin{equation}
    \psi(x) = I(\hat \theta)^{-1/2}\phi(x),
\end{equation}
and is finally used to derive the Fisher kernel, given by
\begin{equation}
    K(x,x')=\psi(x)^T\psi(x').
\end{equation}

\label{sec:kernel}. 
\section{Appendix B}
\label{sec:AB}
We reproduce here some results related to sample size computations for tests of difference of means of multivariate Normal distributions (MVN) in dimension $p$ as can be found in \cite{anderson1962introduction}.

Consider a randomized control trial with two arms. Assume that in the treatment arm, there are $n_T$ iid observations 
\begin{equation}
    X_1,\ldots,X_{n_T} \mbox{ with } X_i \sim MVN(\mu_T,\Sigma)
\end{equation}
In the control arm, there are $n_C$ iid observations
\begin{equation}
    Y_1,\ldots,Y_{n_T} \mbox{ with } Y_i \sim MVN(\mu_C,\Sigma)
\end{equation}
 Note that the covariance is assumed to be the same in each arm, which is a standard simplifying hypothesis. The RCT will test
\begin{equation}
H_0: \mu_T=\mu_C \mbox{ versus }    
\end{equation}
against a local alternative which we define next. As discussed in t5he main text, we assume that we have access to historical data of asymptomatic subjects (A) as well as symptomatic subjects (S). From these two datasets, we compute $\hat \mu_A$ and $\hat \mu_S$, the respective empirical mean. Note that since the dimension $d$ is finite, these are the usual empirical means. We then compute the power at the local alternative
\begin{equation}
H_a:  \mu_T-\mu_{C} = (1-\rho)(\hat \mu_{A}-\hat \mu_{S})  = (1-\rho)d
\end{equation}
The test can then be rephrased as
\begin{equation}
H_0: \rho=0 \mbox{ versus } H_a: \rho = \rho_0    
\end{equation}
A typical value for $\rho_0$ is $40\%$. Note also that we assume that the historical data is sufficiently large compared to the sample size used in the RCT, such that $d$ is considered as a constant and not as random variable. within these assumptions,  
the situation is standard in the statistics litterature, and the results in \cite{anderson1962introduction} are applicable. The statistic to use is the Hotelling T squared statistics. 
\begin{equation}
    T^2=\frac{n_T n_C}{n_T + n_C}(\bar{x}-\bar{y})^T{\hat{\Sigma}}^{-1}(\bar{x}-\bar{y})
\end{equation}
where $\hat{\Sigma}$ is the unbiased pooled covariance matrix estimate
\begin{eqnarray}
\hat{\Sigma} = \frac{(n_T-1)\hat{\Sigma}_x + (n_C-1)\hat{\Sigma}_y}{n_T+n_C-2}
\end{eqnarray}
with 
\begin{equation}
    \bar{x}=\frac{1}{n_T}\sum_{i=1}^{n_T} x_i \mbox{ and }\hat{\Sigma}_x = \frac{1}{n_T-1}\sum_{i=1}^{n_1} (x_i- \bar{x})(x_i-\bar{x})^T
\end{equation}
and similarly for $\bar{y}$ and $\hat{\Sigma}_y$

Moreover, it holds that under $H_0$, 
\begin{equation}
    \frac{n_T+n_C-p-1}{(n_T+n_C-2)p}T^2 \sim F(p,n_T+n_C-1-p)
\end{equation}
where $F$ is the F-distribution. Let
\begin{equation}
    \delta = \frac{n_1 n_2}{n_1 + n_2}\nu^T \hat \Sigma^{-1}\nu
\end{equation}
then, under $H_a$, $T^2$ follows a non-central F-distribution
\begin{equation}
    \frac{n_1 + n_2 -p-1}{(n_1+n_2-2)p}T^2 \sim F(p,n_1+n_2-1-p;\delta) 
\end{equation}
It is common to define the effect size as  
\begin{equation}
    \sqrt{\nu^T \hat \Sigma^{-1}\nu}
\end{equation}
because it provides an expression for the magnitude of the standardized difference
between the group means. 
\section{Appendix C: linear mixed effect model}
\label{sec:AC}
$$y_{i}(t)=\beta_0 + \beta_{i0}+ \beta_{1} t + \beta_{2} \textsf{mci.ind}_i  + \beta_{3} \textsf{mci.ind}_i*t + \varepsilon_{it}.$$ Here $\beta_{0i}$ denotes a subject specific random intercept, $t$ denotes the week number, $\textsf{mci.ind}_i$ represents the indicator variable for the `MCI` group, and $\varepsilon_{it}$ denotes an independent error. We are particularly interested in the power of the coefficient for the interaction term $\textsf{mci.ind}_{i}*t$, which indicates whether or not the two groups differ in their slopes.
\section{Appendix D: Gaussian process model for computer usage data}
\label{sec:AD}
After extensive empirical observation of the computer usage data over time of a collection of elderly normally aging subjects, we propose the following model. We denote $n(t)$ the number of subjects observed at time $t$. We assume that $n(t)>0$ over a range $[0,T]$. We also denote $n(s,t)$ the number of subjects observed both at times $s$ and $t$. We notate $\bar{y}(t)$ the average computer usage for the subjects observed at time $t$. We then propose the following Gaussian Process (GP) model for $\bar{y}(t)$   
\begin{eqnarray}
\bar{y}(t)&=&\bar{y}(0) + \mu t + \bar{w}(t) + \bar{\epsilon}(t), 0 \leq t \leq T\\
 \bar{\epsilon}(t)&=&GP\left(0,\frac{\sigma^2}{n(t)}\delta\right)\\
\bar{w}(t)&=& GP(0,K)\\
K(s,t)&=&\alpha^2\frac{n(s,t)}{n(s)n(t)}s^{\beta/2}t^{\beta/2}e^{-\frac{1}{2\rho^2}|s-t|^\nu}
\end{eqnarray}
We assume that we observe the process $\bar{y}$ at a collection of $m$ time points $t=(t_1,\ldots,t_m)$. The resulting vector
\begin{equation}
    x=\left(\bar{y}(t_1),\ldots,\bar{y}(t_m)\right)
\end{equation}
is then MVN with parameter $\theta=(\theta_1=\mu,\theta_2=\sigma^2,\theta_3=\alpha^2,\theta_4=\beta,\theta_5=\rho^2,\theta_6=\nu)$. These parameter were estimated using the group of asymptomatic subjects. The fitting was obtained using Bayesian modelling and Hamiltonian Monte-Carlo, see \cite{girolami2011riemann}.

In the simulation study, the treated subjects are modelled with the same parameters, aside from the parameter $\mu$, reflecting the effect of the treatment. 

\begin{figure}[h]
\centering
\includegraphics[scale=0.45]{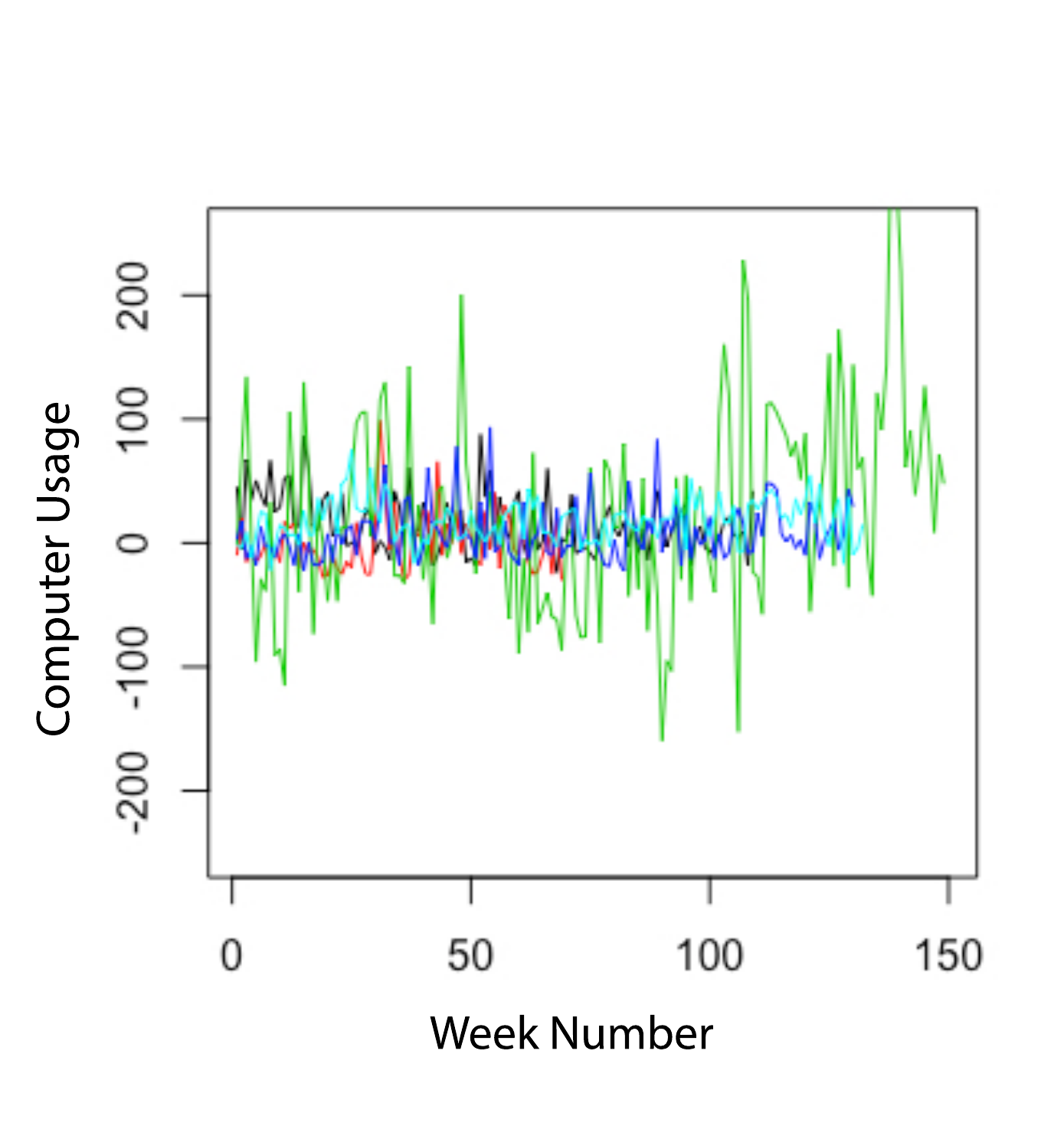}
\includegraphics[scale=0.45]{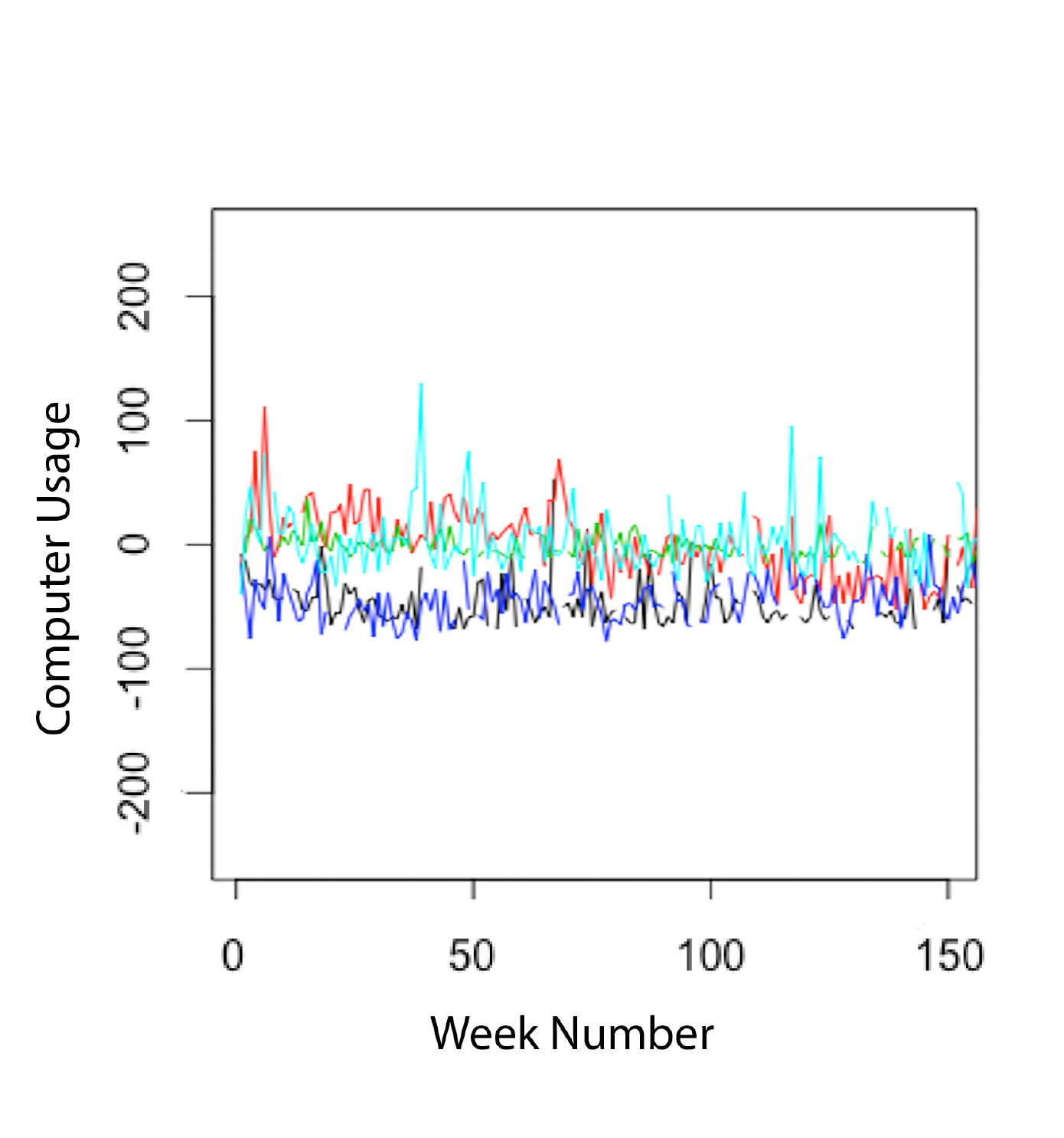}
\caption{Observed computer usage relative to week 0 for 5 subjects. {\bf Left:} Mild Cognitively Impaired  subjects. {\bf Right:} Cognitively normal subjects. }
\label{fig:examples-trajectories}
\end{figure}

\section{Appendix E: preprocessing of the computer usage data fromORCATECH}
\label{sec:AE}
For each subject, we had up to 250 weekly measurements of computer usage. We picked each subjects first week as their first complete week within weeks 5-15 of the study to account for any effects of setting up the monitoring equipment etc. We then selected the subjects next 150 measurements because some subjects were missing data for the remaining 100 weeks. We finally took each measurement and subtracted the measurement from the first week so that each data point is now a change from the subjects first week. During the 150 week period the 11 MCI subjects develop mild cognitive impairment. 

\bibliographystyle{unsrt}
\bibliography{neurips_2020}

\begin{thebibliography}{10}

\bibitem{vos2016global}
Theo Vos, Christine Allen, Megha Arora, Ryan~M Barber, Zulfiqar~A Bhutta,
  Alexandria Brown, Austin Carter, Daniel~C Casey, Fiona~J Charlson, Alan~Z
  Chen, et~al.
\newblock Global, regional, and national incidence, prevalence, and years lived
  with disability for 310 diseases and injuries, 1990--2015: a systematic
  analysis for the global burden of disease study 2015.
\newblock {\em The lancet}, 388(10053):1545--1602, 2016.

\bibitem{favela2015living}
Jesus Favela, Jeffrey Kaye, Marjorie Skubic, Marilyn Rantz, and Monica Tentori.
\newblock Living labs for pervasive healthcare research.
\newblock {\em IEEE Pervasive Computing}, 14(2):86--89, 2015.

\bibitem{dodge2015web}
Hiroko~H Dodge, Jian Zhu, Nora~C Mattek, Molly Bowman, Oscar Ybarra,
  Katherine~V Wild, David~A Loewenstein, and Jeffrey~A Kaye.
\newblock Web-enabled conversational interactions as a method to improve
  cognitive functions: Results of a 6-week randomized controlled trial.
\newblock {\em Alzheimer's \& Dementia: Translational Research \& Clinical
  Interventions}, 1(1):1--12, 2015.

\bibitem{scholkopf2002learning}
Bernhard Sch{\"o}lkopf, Alexander~J Smola, Francis Bach, et~al.
\newblock {\em Learning with kernels: support vector machines, regularization,
  optimization, and beyond}.
\newblock MIT press, 2002.

\bibitem{kung2014kernel}
Sun~Yuan Kung.
\newblock {\em Kernel methods and machine learning}.
\newblock Cambridge University Press, 2014.

\bibitem{gretton2012kernel}
Arthur Gretton, Karsten~M Borgwardt, Malte~J Rasch, Bernhard Sch{\"o}lkopf, and
  Alexander Smola.
\newblock A kernel two-sample test.
\newblock {\em The Journal of Machine Learning Research}, 13(1):723--773, 2012.

\bibitem{eric2008testing}
Moulines Eric, Francis~R Bach, and Za{\"\i}d Harchaoui.
\newblock Testing for homogeneity with kernel fisher discriminant analysis.
\newblock In {\em Advances in Neural Information Processing Systems}, pages
  609--616, 2008.

\bibitem{harchaoui2013kernel}
Zaid Harchaoui, Francis Bach, Olivier Cappe, and Eric Moulines.
\newblock Kernel-based methods for hypothesis testing: A unified view.
\newblock {\em IEEE Signal Processing Magazine}, 30(4):87--97, 2013.

\bibitem{fukumizu2009kernel}
Kenji Fukumizu, Arthur Gretton, Gert~R Lanckriet, Bernhard Sch{\"o}lkopf, and
  Bharath~K Sriperumbudur.
\newblock Kernel choice and classifiability for rkhs embeddings of probability
  distributions.
\newblock In {\em Advances in neural information processing systems}, pages
  1750--1758, 2009.

\bibitem{li2020adaptable}
Haoran Li, Alexander Aue, Debashis Paul, Jie Peng, Pei Wang, et~al.
\newblock An adaptable generalization of hotelling’s $t^2$ test in high
  dimension.
\newblock {\em Annals of Statistics}, 48(3):1815--1847, 2020.

\bibitem{chen2011regularized}
Lin~S Chen, Debashis Paul, Ross~L Prentice, and Pei Wang.
\newblock A regularized hotelling’s t 2 test for pathway analysis in
  proteomic studies.
\newblock {\em Journal of the American Statistical Association},
  106(496):1345--1360, 2011.

\bibitem{goodfellow2016deep}
Ian Goodfellow, Yoshua Bengio, Aaron Courville, and Yoshua Bengio.
\newblock {\em Deep learning}, volume~1.
\newblock MIT press Cambridge, 2016.

\bibitem{doersch2016tutorial}
Carl Doersch.
\newblock Tutorial on variational autoencoders.
\newblock {\em arXiv preprint arXiv:1606.05908}, 2016.

\bibitem{goodfellow2014generative}
Ian Goodfellow, Jean Pouget-Abadie, Mehdi Mirza, Bing Xu, David Warde-Farley,
  Sherjil Ozair, Aaron Courville, and Yoshua Bengio.
\newblock Generative adversarial nets.
\newblock In {\em Advances in neural information processing systems}, pages
  2672--2680, 2014.

\bibitem{anderson1962introduction}
Theodore~Wilbur Anderson.
\newblock An introduction to multivariate statistical analysis.
\newblock Technical report, Wiley New York, 1962.

\bibitem{girolami2011riemann}
Mark Girolami and Ben Calderhead.
\newblock Riemann manifold langevin and hamiltonian monte carlo methods.
\newblock {\em Journal of the Royal Statistical Society: Series B (Statistical
  Methodology)}, 73(2):123--214, 2011.

\end{thebibliography}

\end{document}